\documentclass[preprint,showpacs,aps]{revtex4}
\newcommand{\UQ}{School of Mathematics and Physics, University of Queensland, Brisbane, 
QLD 4072, Australia.}

\usepackage{graphicx,psfrag}
\usepackage{bm}

\newcommand{\etal}{{\em et al.}}

\newcommand{\pr}{Phys. Rev. }
\newcommand{\jpb}{J. Phys. B }

\newcommand{\jpa}{J. Phys. A }

\newcommand{\ajp}{Am. J. Phys. }
\newcommand{\njp}{New. J. Phys. }

\begin{document}
\title{Tripartite and bipartite entanglement in continuous-variable tripartite systems}

\author{M.~K. Olsen and J.~F. Corney}
\affiliation{\UQ}
\date{\today}

\begin{abstract}

We examine one asymmetric adnd two fully symmetric Gaussian continuous-variable systems in terms of their tripartite and bipartite entanglement properties. We treat pure states and are able to find analytic solutions using the undepleted pump approximation for the Hamiltonian models, and standard beamsplitter relations for a model that mixes the outputs of optical parametric oscillators. Our two symmetric systems exhibit perfect tripartite correlations, but only in the unphysical limit of infinite squeezing. For more realistic squeezing parameters, all three systems exhibit both tripartite and bipartite entanglement. We conclude that none of the outputs are completely analogous to either GHZ or W states, but there are parameter regions where they produce T states introduced by Adesso \etal The qualitative differences in the output states for different interaction parameters indicate that continuous-variable tripartite quantum information systems offer a versatility not found in bipartite systems.  

\end{abstract}


\maketitle

\section{Introduction}
\label{sec:intro}

In the theory of discrete variable tripartite entanglement, the two most famous states are known as the GHZ (Greenberger-Horne-Zeilinger) state~\cite{GHSZ} and the W state~\cite{Guifre}. One essential difference between these two is seen when one mode is traced over and any resulting bipartite entanglement is looked for. The reduced GHZ state then becomes completely separable, while the reduced W state will demonstrate remnant bipartite entanglement. The concept of these two states when transferred to the Gaussian continuous-variable (CV) quantum information~\cite{Braunstein} has been extensively analysed by Adesso \etal~\cite{Adesso1,Adesso2}, who also introduced the T states, which exhibit tripartite entanglement only. 

In this paper we consider three processes which are known to produce CV entanglement and show analytically that they also produce bipartite entanglement in some, but not all, of the parameter space. We detect the bipartite entanglement using the Duan-Simon measure~\cite{Duan,Simon}, which is both necessary and sufficient for Gaussian systems. The fact that both bipartite and tripartite entanglement are exhibited suggests that they are none of the above states over the entire operating regime. Since the bipartite entanglement is not maximal, they are not proper W states, and because the bipartite entanglement exists, they are neither GHZ nor T states. There are also operating regimes where only tripartite entanglement exists, although this can only be perfect in the limit of infinite squeezing. As this limit is not physical, they necessarily lack one of the characteristics of the true GHZ state. Since no bipartite entanglement exists in these regions, the outputs then qualify as T states. 
  
We write the Hamiltonians in the non-depleted pump approximations for the two nonlinear systems~\cite{Smithers,Pfister} and use standard beamsplitter relations for the other~\cite{Aoki}. The solutions in terms of  expectation values of the second moments of the quadrature operators allow us to find analytical expressions for the Duan-Simon~\cite{Duan,Simon} and Reid Einstein-Podolsky-Rosen (EPR)~\cite{EPR,eprMDR} correlations used to denote bipartite entanglement and EPR-steering~\cite{Wisesteer}. We also calculate the correlations developed by van Loock and Furusawa~\cite{VLF} for the detection of tripartite entanglement and the three-mode EPR-steering correlations developed by Olsen \etal~\cite{EPR3}. 

Our results, showing that continuous-variable systems for the production of tripartite entanglement will behave qualitatively differently as the input fields are varied, add a dimension to continuous-variable quantum information which is not present in the discrete variable version. We will show that, by simple tuning of the inputs, different classes of output states are produced. This tunability brings a versatility to these systems that is not present in the discrete variable equivalents, and opens new possibilities for quantum information technologies such as quantum key distribution~\cite{1SQKD}. 

 \section{Common classes of tripartite state}
 \label{sec:classes}
 
In the discrete variable regime, there are two classes of tripartite entangled states of three qubits that are commonly considered. The first of these is the GHZ state, introduced by Greenberger \etal~\cite{GHSZ}, commonly represented as
\begin{equation}
|\psi_{GHZ}\rangle = \frac{1}{\sqrt{2}}\left(|000\rangle + |111\rangle \right).
\label{eq:GHZdef}
\end{equation}
This state possesses maximal tripartite entanglement and no remnant bipartite entanglement whatsoever whenever any one of the qubits is traced over. This state also gives a maximal Bell violation.

The canonical representative of the W states is written as
\begin{equation}
|\psi_{W}\rangle = \frac{1}{\sqrt{3}}\left(|100\rangle + |010\rangle + |001\rangle \right),
\label{eq:Wdef}
\end{equation} 
which also exhibits maximal tripartite entanglement but also exhibits remnant bipartite entanglement if any one mode is traced over~\cite{Guifre}. Both of these states as written are pure and symmetric.

When we consider continuous-variable tripartite states, the situation becomes somewhat different. The systems demonstrated by Smithers~\cite{Smithers} and Aoki~\cite{Aoki} will exhibit the perfect correlations of a GHZ state in the regime where they become perfect quadrature eigenstates. Unfortunately, this limit is not obtainable in practice. In the Aoki scheme, for example, it would require perfect squeezing, which is unphysical. In the version of the Smithers scheme that we will analyse here, it would require infinite interaction strength, which is also not a physically relevant concept. However, Adesso and Illuminati~\cite{Adesso1} have addressed this problem, dividing CV tripartite states into five separate classes. These range from states which are totally inseparable under any of the three possible bipartitions to those which are separable under all possible bipartitions.  In terms of the entanglement properties, they add a type to the GHZ-types and W-types commonly in use. For a state which exhibits tripartite entanglement only, but without being in the GHZ limit, they introduce the nomenclature T state. We will use this nomenclature in this article.

\section{The inequalities}
\label{sec:inequalities}

\subsection{Bipartite measures}
\label{subsec:Bipart}

We will evaluate bipartite entanglement and EPR-steering in terms of the functions of quadrature operators developed by Duan \etal~\cite{Duan}, Simon~\cite{Simon}, and Reid~\cite{eprMDR}, which are appropriate measures for two-mode optical systems.
The quadrature operators are defined as $\hat{X}_{i}=\hat{a}_{i}+\hat{a}^{\dag}_{i}$ and $\hat{Y}_{i}=-i(\hat{a}_{i}-\hat{a}^{\dag}_{i})$. This allows us to define the Duan-Simon inequalities as
\begin{eqnarray}
V(\hat{X}_{i}+\hat{X}_{j})+V(\hat{Y}_{i}-\hat{Y}_{j}) &\geq& 4, \nonumber \\
V(\hat{X}_{i}-\hat{X}_{j})+V(\hat{Y}_{i}+\hat{Y}_{j}) &\geq& 4,
\label{eq:DSinequalities}
\end{eqnarray}
with $i$ and $j$ being mode indices. Violation of either of these is a demonstration of bipartite entanglement. We will call the first of these combined variance sums $DS_{ij}^{+}$ and the second $DS_{ij}^{-}$. Because all the systems we consider are Gaussian and pure, these entanglement correlations are both necessary and sufficient for this demonstration~\cite{Teh}.

The EPR paradox is detected by the well-known criteria developed by Reid~\cite{eprMDR}, in terms of inferred quadrature variances,
\begin{equation}
V^{inf}(\hat{X}_{i})V^{inf}(\hat{Y}_{i})<1.
\label{eq:eprMDR}
\end{equation}
The inferred variances are defined, with the value of $\hat{X}_{i}$ being inferred from measurements of $\hat{X}_{j}$ (and similarly for $\hat{Y}_{i}$), as,
\begin{eqnarray}
V^{inf}(\hat{X}_{i}) &=& V(\hat{X}_{i}) - \frac{\left[V(\hat{X}_{i},\hat{X}_{j}\right]^{2}}{V(\hat{X}_{j})},\nonumber\\
V^{inf}(\hat{Y}_{i}) &=& V(\hat{Y}_{i}) - \frac{\left[V(\hat{Y}_{i},\hat{Y}_{j}\right]^{2}}{V(\hat{Y}_{j})},
\label{eq:MDRinfs}
\end{eqnarray}
from which we immediately see that there is an implied asymmetry since we can equally define $V^{inf}(\hat{X}_{j})$, swapping the roles of the people measuring each mode. In some circumstances the values measured at $i$ can be inferred from measurements on mode $j$, but not vice-versa. This was first predicted, in sum frequency generation, by Olsen and Bradley~\cite{SFG} and has recently been further analysed by Ji \etal~\cite{Korea}. In what follows we will label the product $V^{inf}(\hat{X}_{i})V^{inf}(\hat{Y}_{i})$, inferred from $\hat{X}_{j}$ and $\hat{Y}_{j}$, as $\Pi V_{ij}$.

\subsection{Tripartite}
\label{subsec:tripart}

The van Loock-Furusawa conditions~\cite{VLF} give a set of inequalities
\begin{eqnarray}
V_{12} &=& V(\hat{X}_{1}-\hat{X}_{2})+V(\hat{Y}_{1}+\hat{Y}_{2}+g_{3}\hat{Y}_{3}) \geq 4, \nonumber \\
V_{13} &=& V(\hat{X}_{1}-\hat{X}_{3})+V(\hat{Y}_{1}+g_{2}\hat{Y}_{2}+\hat{Y}_{3}) \geq 4, \nonumber \\
V_{23} &=& V(\hat{X}_{2}-\hat{X}_{3})+V(g_{1}\hat{Y}_{1}+\hat{Y}_{2}+\hat{Y}_{3}) \geq 4,
\label{eq:VLF}
\end{eqnarray}
for which the violation of any two demonstrates tripartite entanglement. The $g_{j}$, which are arbitrary and real, can be optimised~\cite{AxMuzz}, using the variances and covariances, as
\begin{eqnarray}
g_{1} &=& -\frac{V(\hat{Y}_{1},\hat{Y}_{2})+V(\hat{Y}_{1},\hat{Y}_{3})}{V(\hat{Y}_{1})}, \nonumber \\
g_{2} &=& -\frac{V(\hat{Y}_{1},\hat{Y}_{2})+V(\hat{Y}_{2},\hat{Y}_{3})}{V(\hat{Y}_{2})}, \nonumber \\
g_{3} &=& -\frac{V(\hat{Y}_{1},\hat{Y}_{3})+V(\hat{Y}_{2},\hat{Y}_{3})}{V(\hat{Y}_{3})},
\label{eq:VLFopt}
\end{eqnarray}
which is the process we follow with the results presented below.

Another set of inequalities was also presented by van Loock and Furusawa, the violation of any one of which is sufficient to prove tripartite entanglement,
\begin{eqnarray}
V_{123} &=& V(\hat{X}_{1}-\frac{\hat{X}_{2}+\hat{X}_{3}}{\sqrt{2}})+V(\hat{Y}_{1}+\frac{\hat{Y}_{2}+\hat{Y}_{3}}{\sqrt{2}}) \geq 4,\nonumber \\
V_{312} &=& V(\hat{X}_{3}-\frac{\hat{X}_{1}+\hat{X}_{2}}{\sqrt{2}})+V(\hat{Y}_{1}+\frac{\hat{Y}_{1}+\hat{Y}_{2}}{\sqrt{2}}) \geq 4,\nonumber \\
V_{231} &=& V(\hat{X}_{2}-\frac{\hat{X}_{1}+\hat{X}_{3}}{\sqrt{2}})+V(\hat{Y}_{2}+\frac{\hat{Y}_{1}+\hat{Y}_{3}}{\sqrt{2}}) \geq 4.
\label{eq:VLFijk}
\end{eqnarray}

The generalisation of the Reid EPR inequalities to three modes by Olsen \etal~\cite{EPR3} involves using either one mode to infer combined properties of the other two, or combined properties of two of the modes to infer properties of the third mode. We define
\begin{eqnarray}
V^{inf}(\hat{X}_{i}) &=& V(\hat{X}_{i})-\frac{[V(\hat{X}_{i},\hat{X}_{j}\pm\hat{X}_{k}]^{2}}{V(\hat{X}_{j}\pm\hat{X}_{k})}, \nonumber \\
V^{inf}(\hat{Y}_{i}) &=& V(\hat{Y}_{i})-\frac{[V(\hat{Y}_{i},\hat{Y}_{j}\pm\hat{Y}_{k}]^{2}}{V(\hat{Y}_{j}\pm\hat{Y}_{k})},
\label{eq:EPRi} 
\end{eqnarray}
with a demonstration of the paradox requiring
\begin{equation}
V^{inf}(\hat{X}_{i})V^{inf}(\hat{Y}_{i}) < 1.
\label{eq:VXiVYi}
\end{equation}
When this is satisfied for $i$, $j$ and $k$, we have established tripartite entanglement. We can also use the inferred variances of the combined modes
\begin{eqnarray}
V^{inf}(\hat{X}_{j}\pm\hat{X}_{k}) &=& V(\hat{X}_{j}\pm\hat{X}_{k})-\frac{[V(\hat{X}_{i},\hat{X}_{j})\pm V(\hat{X}_{i},\hat{X}_{k})]^{2}}{V(\hat{X}_{i})}, \nonumber \\
V^{inf}(\hat{Y}_{j}\pm\hat{Y}_{k}) &=& V(\hat{Y}_{j}\pm\hat{Y}_{k})-\frac{[V(\hat{Y}_{i},\hat{Y}_{j})\pm V(\hat{Y}_{i},\hat{Y}_{k})]^{2}}{V(\hat{Y}_{i})},
\label{eq:VXjkVYjk} 
\end{eqnarray}
with a demonstration of the paradox when
\begin{equation}
V^{inf}(\hat{X}_{j}\pm\hat{X}_{k})V^{inf}(\hat{Y}_{j}\pm\hat{Y}_{k}) < 4.
\label{eq:EPR3jk}
\end{equation} 
In the interests of brevity we will label these two correlations $\Pi^{(3)}V_{i}$ (Eq.~\ref{eq:VXiVYi}) and  $\Pi^{(3)}V_{ij}$ (Eq.~\ref{eq:EPR3jk}).    
As above, a demonstration for the three possible combinations establishes tripartite entanglement. In the language of EPR-steering introduced by Wiseman \etal~\cite{Wisesteer}, a demonstration via $\Pi^{(3)}V_{i}$ means that two of the participants have combined to steer the third. A demonstration via $\Pi^{(3)}V_{ij}$ means that one participant can steer the combined properties measured by the other two, without steering either of them individually.

\section{A symmetric model from a single Optical Parametric Amplifier}
\label{sec:Pfister}

This model consists of triply concurrent downconversion~\cite{Olivier}, with the intracavity version being analysed by Bradley \etal~\cite{Pfister}, where it was noted that the state created tended towards a GHZ state in the limit of infinite squeezing, but was analogous to a W state for finite squeezing.  The interaction Hamiltonian in the undepleted pump approximation is written as
\begin{equation}
{\cal H}_{int} = i\hbar\kappa\left[\hat{a}_{1}^{\dag}\hat{a}_{2}^{\dag}+\hat{a}_{1}^{\dag}\hat{a}_{3}^{\dag}+\hat{a}_{2}^{\dag}\hat{a}_{3}^{\dag}-
\hat{a}_{1}\hat{a}_{2}-\hat{a}_{1}\hat{a}_{3}-\hat{a}_{2}\hat{a}_{3}\right] .
\end{equation}
where $\kappa$ represents the product of the optical nonlinearity and the corresponding pump fields. Note that we have set the two pump fields as equal. 
Setting
\begin{eqnarray}
A &=& \cosh 2\kappa t+2 \cosh\kappa t, \nonumber \\
B &=& \sinh 2\kappa t-2\sinh\kappa t, \nonumber \\
C &=& \cosh 2\kappa t - \cosh\kappa t,\nonumber \\
D &=& \sinh\kappa t+ \sinh 2\kappa t,
\label{eq:ABCD} 
\end{eqnarray}
the Heisenberg equations of motion for the quadrature operators are solved as~\cite{EPR3}
\begin{eqnarray}
\hat{X}_{1}(t) &=& \frac{1}{3}\left[(A+B)\hat{X}_{1}(0)+(C+D)\hat{X}_{2}(0)+(C+D)\hat{X}_{3}(0) \right], \nonumber \\
\hat{X}_{2}(t) &=& \frac{1}{3}\left[(C+D)\hat{X}_{1}(0)+(A+B)\hat{X}_{2}(0)+(C+D)\hat{X}_{3}(0) \right], \nonumber \\
\hat{X}_{3}(t) &=& \frac{1}{3}\left[(C+D)\hat{X}_{1}(0)+(C+D)\hat{X}_{2}(0)+(A+B)\hat{X}_{3}(0) \right], \nonumber \\
\hat{Y}_{1}(t) &=& \frac{1}{3}\left[(A-B)\hat{Y}_{1}(0)+(C-D)\hat{Y}_{2}(0)+(C-D)\hat{Y}_{3}(0) \right], \nonumber \\
\hat{Y}_{2}(t) &=& \frac{1}{3}\left[(C-D)\hat{Y}_{1}(0)+(A-B)\hat{Y}_{2}(0)+(C-D)\hat{Y}_{3}(0) \right], \nonumber \\
\hat{Y}_{3}(t) &=& \frac{1}{3}\left[(C-D)\hat{Y}_{1}(0)+(C-D)\hat{Y}_{2}(0)+(A-B)\hat{Y}_{3}(0) \right],
\label{eq:HPXY}
\end{eqnarray}
from which we can calculate the second order moments necessary for the Duan-Simon, Reid EPR, and van Loock-Furusawa (VLF) correlations.
We find
\begin{eqnarray}
\langle\hat{X}_{i}^{2}(t)\rangle &=& \frac{1}{9}\left[(A+B)^{2}+2(C+D)^{2} \right], \nonumber \\
\langle\hat{Y}_{i}^{2}(t)\rangle &=& \frac{1}{9}\left[(A-B)^{2}+2(C-D)^{2} \right], \nonumber \\
\langle\hat{X}_{i}\hat{X}_{j}(t)\rangle &=& \frac{1}{9}\left[(C+D)(2A+2B+C+D) \right], \nonumber \\
\langle\hat{Y}_{i}\hat{Y}_{j}(t)\rangle &=& \frac{1}{9}\left[(C-D)(2A+C-2B-D) \right],
\label{eq:HPXXYY}
\end{eqnarray}
all of which assume vacuum in these modes at $t=0$.

\begin{figure}[tbp]
\includegraphics[width=0.75\columnwidth]{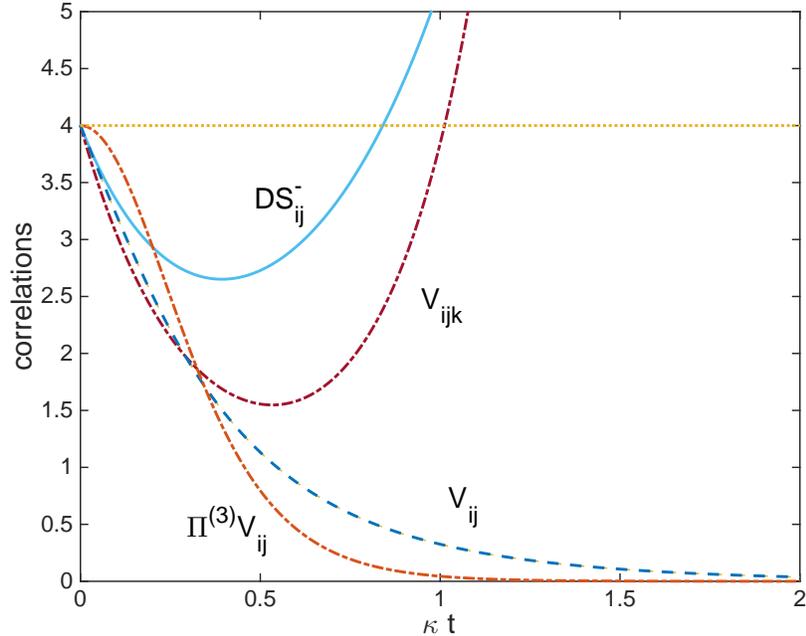}
\caption{(colour online) DS$_{ij}^{-}$, V$_{ij}$, V$_{ijk}$, and $\Pi^{3}$V$_{ij}$ for the single OPA model. Note that these correlations do not change under permutations of the indices. The line at $4$ is a guide to the eye. In this and all subsequent graphs, the results are dimensionless.}
\label{fig:Pfisterfig}
\end{figure}

In Fig.~\ref{fig:Pfisterfig} we show the results for the various correlations. Interestingly, despite the fact that we know this system displays full tripartite entanglement, we see that the DS$_{ij}^{-}$ detect a degree of bipartite entanglement at early times. If the state produced were analogous to a continuous-variable GHZ state, there would be no possibility of bipartite entanglement~\cite{Guifre}. In fact, if we operate on the three-mode vacuum state with the interaction Hamiltonian, we find to first order
\begin{equation}
\psi(\Delta t) \approx |0,0,0\rangle + \frac{\kappa\Delta t}{\sqrt{3}}\left(|1,1,0\rangle+|1,0,1\rangle+|0,1,1\rangle \right),
\label{eq:almostW}
\end{equation}
which has more in common with the W states~\cite{Guifre,Fortescue},
so that a degree of bipartite entanglement is therefore not ruled out. In fact we can see that tracing over any one mode leaves bipartite entanglement for both $\psi(\Delta t)$ and $|\psi_{W}\rangle$. We found no evidence of bipartite EPR-steering in this system, with the correlation $\Pi V_{ij}$ never being less than one. 
Fig.~\ref{fig:Pfisterfig} also shows the results for the two van Loock-Furusawa correlations and the EPR-steering correlation $\Pi^{(3)}V_{ij}$. All of these are equal under any exchange of indices for this system and $\Pi^{(3)}V_{ijk}$ is an identical shape to $\Pi^{(3)}V_{ij}$, but begins at a value of one. We see that the system begins to exhibit tripartite entanglement and EPR-steering as soon as it is turned on, but that the correlation $V_{ijk}$ fails to detect this after some time. In the region where DS$_{ij}^{-}>4$, the output is properly caled a T state~\cite{Adesso1,Adesso2}. It becomes more analogous to a GHZ state (which would require $V_{ij}=0$) as $\kappa t$ increases, but is only totally equivalent in the unphysical limit of infinite interaction time $\kappa t$.

\section{Three OPOs and a two beamsplitters}
\label{sec:BSOPO}

One of the pioneering results for continuous variable tripartite entanglement came from van Loock and Braunstein~\cite{VLB}, and was implemented experimentally by
Aoki \etal~\cite{Aoki}, who mixed three squeezed states on two beamsplitters to obtain three entangled output beams, as shown in
This setup was subsequently analysed in terms of tripartite entanglement and EPR-steering in both the time and frequency domains by Olsen \etal~\cite{EPR3} and is a subset of the systems recently analysed by Wang \etal~\cite{WangGHZ}. Bipartite measures were not analysed in these articles.
The system uses three optical parametric oscillators (OPO), with the first, OPO$_{1}$, producing a state squeezed in the $\hat{Y}$ quadrature, while the other two produce $\hat{X}$ squeezed states. The annihilation operator $\hat{a}_{j}$ represents the output of OPO$_{j}$. The output of OPO$_{1}$ and OPO$_{2}$ are mixed on the first beamsplitter, BS$_{1}$, to produce outputs represented by $\hat{b}_{0}$ and $\hat{b}_{1}$. The field corresponding to $\hat{b}_{0}$  is then mixed with $\hat{a}_{3}$ on BS$_{2}$. The outputs of BS$_{2}$ are represented by $\hat{b}_{2}$ and $\hat{b}_{3}$. With the squeezed inputs, tripartite entanglement is found between the three outputs. In Ref.~\cite{EPR3}, the van Loock Furusawa correlations $V_{ij}$ were calculated analytically, but without optimisation. The tripartite EPR-steering correlations were also calculated. We will now calculate the $V_{ij}$ correlations with optimisation, and examine both bipartite and tripartite entanglement in this system.

 Assigning BS$_{1}$ a reflectivity of $\mu$ and BS$_{2}$ a reflectivity of $\nu$, we find the solutions for the $\hat{b}_{j}$ in terms of the inputs as 
 \begin{eqnarray}
 \hat{b}_{1} &=& \sqrt{1-\mu}\;\hat{a}_{1}+\sqrt{\mu}\;\hat{a}_{2}, \nonumber \\
 \hat{b}_{2} &=& \sqrt{\mu(1-\nu)}\;\hat{a}_{1}-\sqrt{(1-\mu)(1-\nu)}\;\hat{a}_{2}+\sqrt{\nu}\;\hat{a}_{3}, \nonumber \\
 \hat{b}_{3} &=& \sqrt{\mu\nu}\;\hat{a}_{1}-\sqrt{\nu(1-\mu)}\;\hat{a}_{2}-\sqrt{1-\nu}\;\hat{a}_{3},
\label{eq:BSErnie} 
\end{eqnarray}
which allow us to find all the correlations we require for the bipartite and tripartite correlations we wish to calculate. We note here that, although these expressions appear asymmetric, they become fully symmetric for $\mu=2/3$ and $\nu=1/2$, and these are the values we use in our final results. The required variances are
\begin{eqnarray}
V(\hat{X}_{b_{1}}) &=& (1-\mu)V(\hat{X}_{a_{1}})+\mu V(\hat{X}_{a_{2}}), \nonumber \\
V(\hat{Y}_{b_{1}}) &=& (1-\mu) V(\hat{Y}_{a_{1}})+\mu V(\hat{Y}_{a_{2}}), \nonumber \\
V(\hat{X}_{b_{2}}) &=& \mu(1-\nu) V(\hat{X}_{a_{1}})+(1-\nu)(1-\mu) V(\hat{X}_{a_{2}})+\nu V(\hat{X}_{a_{3}}), \nonumber \\
V(\hat{Y}_{b_{2}}) &=& \mu(1-\nu) V(\hat{Y}_{a_{1}})+(1-\nu)(1-\mu) V(\hat{Y}_{a_{2}})+\nu V(\hat{Y}_{a_{3}}), \nonumber \\
V(\hat{X}_{b_{3}}) &=& \mu\nu V(\hat{X}_{a_{1}}) + \nu(1-\mu)V(\hat{X}_{a_{2}})+(1-\nu)V(\hat{X}_{a_{3}}), \nonumber \\
V(\hat{Y}_{b_{3}}) &=& \mu\nu V(\hat{Y}_{a_{1}}) + \nu(1-\mu)V(\hat{Y}_{a_{2}})+(1-\nu)V(\hat{Y}_{a_{3}}),
\label{eq:BSVXY}
\end{eqnarray}
and the covariances are
\begin{eqnarray}
V(\hat{X}_{b_{1}},\hat{X}_{b_{2}}) &=& \sqrt{\mu(1-\mu)(1-\nu)}\left[V(\hat{X}_{a_{1}})-V(\hat{X}_{a_{2}}) \right], \nonumber \\
V(\hat{X}_{b_{1}},\hat{X}_{b_{3}}) &=& \sqrt{\mu\nu(1-\mu)}\left[V(\hat{X}_{a_{1}})-V(\hat{X}_{a_{2}}) \right], \nonumber \\
V(\hat{X}_{b_{2}},\hat{X}_{b_{3}}) &=& \sqrt{\nu(1-\nu)}\left[\mu V(\hat{X}_{a_{1}})+(1-\mu)V(\hat{X}_{a_{2}})-V(\hat{X}_{a_{3}}) \right], \nonumber \\
V(\hat{Y}_{b_{1}},\hat{Y}_{b_{2}}) &=& \sqrt{\mu(1-\mu)(1-\nu)}\left[V(\hat{Y}_{a_{1}})-V(\hat{Y}_{a_{2}}) \right], \nonumber \\
V(\hat{Y}_{b_{1}},\hat{Y}_{b_{3}}) &=& \sqrt{\mu\nu(1-\mu)}\left[V(\hat{Y}_{a_{1}})-V(\hat{Y}_{a_{2}}) \right], \nonumber \\
V(\hat{Y}_{b_{2}},\hat{Y}_{b_{3}}) &=& \sqrt{\nu(1-\nu)}\left[\mu V(\hat{Y}_{a_{1}})+(1-\mu)V(\hat{Y}_{a_{2}})-V(\hat{Y}_{a_{3}}) \right],
\label{eq:BSCovs}
\end{eqnarray}
from which we have all that is necessary to calculate the Duan-Simon, van Loock Furusawa and EPR-steering correlations. For $\mu=2/3$ and $\nu=1/2$ as in Aoki \etal~\cite{Aoki}, the $V_{ij}$ and three-mode EPR correlations are given in Ref.~\cite{EPR3}. However, possible bipartite entanglement was not analysed in that work, nor were the $V_{ij}$ optimised using the $g_{i}$, so we will give these results here.

For a squeezing parameter $r$, equal for each OPO, we may assume minimum uncertainty squeezed states and set 
\begin{eqnarray}
V(\hat{X}_{a_{1}}) &=& V(\hat{Y}_{a_{2}}) = V(\hat{Y}_{a_{3}}) = \mbox{e}^{r}, \nonumber \\
V(\hat{Y}_{a_{1}}) &=& V(\hat{X}_{a_{2}}) = V(\hat{X}_{a_{3}}) = \mbox{e}^{-r},
\label{eq:minimumuncertainty}
\end{eqnarray}
which leads to the bipartite correlations
\begin{equation}
DS_{ij}^{\pm} = 4\cosh r \pm \frac{8}{3}\sinh r,
\label{eq:DSDS}
\end{equation}
of which $DS_{ij}^{-}$ falls below $4$ over a range of $r$. For these parameters, we do not see a demonstration of bipartite EPR-steering.

The optimised $V_{ij}$ are found as
\begin{equation}
V_{ij} = \frac{2+10\mbox{e}^{2r}}{\mbox{e}^{r}+2\mbox{e}^{3r}},
\label{eq:VijAoki}
\end{equation}
and the $V_{ijk}$ are
\begin{equation}
V_{ijk} = 4\left(\cosh r-\frac{2\sqrt{2}}{3}\sinh r \right),
\label{eq:VijkAoki}
\end{equation}
with these not changing under permutations of the indices. Note that, with optimisation, the $V_{ij}$ begin at $4$, rather than at the non-optimised value of $5$ found in Ref.~\cite{EPR3}. For completeness we note that
\begin{eqnarray}
\Pi^{(3)}V_{i} &=& \frac{9}{5+4\cosh 2r}, \nonumber\\
\Pi^{(3)}V_{ij} &=& \frac{36}{5+4\cosh 2r},
\label{eq:AokiEPR3}
\end{eqnarray}
so that the two types of tripartite EPR-steering become available as soon as $r$ is greater than zero.

\begin{figure}[tbp]
\includegraphics[width=0.75\columnwidth]{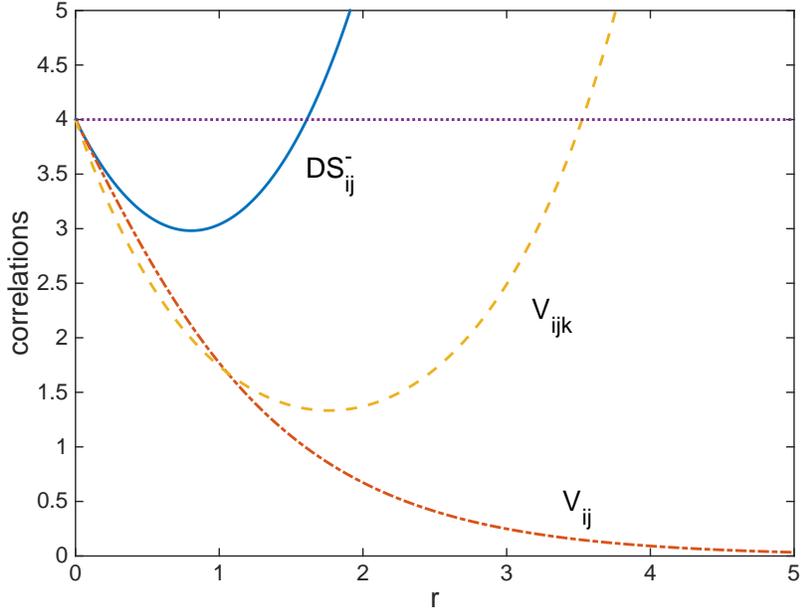}
\caption{(colour online) The DS$^{-}_{ij}$, V$_{ij}$ and V$_{ijk}$ correlations for the Aoki scheme with $\mu=2/3$ and $\nu=1/2$, showing that bipartite and tripartite entanglement are both available over a range of squeezing. Note that the line at $4$ is a guide to the eye. The dimensionless squeezing parameter is denoted by $r$.}
\label{fig:BSbitri}
\end{figure}

We see from Fig.~\ref{fig:BSbitri} that the V$_{ijk}$ once again fail to detect entanglement in a parameter regime where it is found by the V$_{ij}$. 
In the Aoki experiment~\cite{Aoki}, a V$_{ij}$ of approximately $3$ was measured. The fact that we found no bipartite EPR-steering is consistent with the result of Wang~\etal~\cite{WangGHZ}, who showed that, in an $N$-mode system of this type, with $N-1$ beamsplitters and $N$ sources, at least $N/2$ participants must combine to steer any single participant. Fig.~\ref{fig:BSbitri} shows that the experimental result of Ref.~\cite{Aoki} had not entered the T state regime, which begins for V$_{ij}\approx 1$. In the experiment, both bipartite and tripartite entanglement would have been available, with an increase in squeezing needed for the bipartite entanglement to disappear. As with the scheme of section~\ref{sec:Pfister}, the system tends towards a state with true GHZ properties only in the limit of large squeezing. In fact, the behaviours of the two systems in terms of quantum correlations are very similar, which is as expected since both are examples of fully symmetric Gaussian systems.

\section{An asymmetric model}
\label{sec:Clervie}

Our asymmetric system, which combines downconversion with sum-frequency generation, was first proposed by Smithers and  Lu~\cite{Smithers}, and theoretically analysed in both travelling wave~\cite{Ferraro} and in an intracavity configuration by Yu \etal~\cite{Yu}. The configuration was subsequently analysed in more depth by Pennarun \etal~\cite{Clervie}, who investigated the stability properties and predicted tripartite entanglement in different regimes. It consists of a nonlinear medium pumped at frequency $\omega_{0}$. The downconversion part of the process, denoted by the effective nonlinearity $\kappa_{1}$, generates two fields at  
$\omega_{1}$ and $\omega_{3}$, where $\omega_{0}=\omega_{1}+\omega_{3}$. The pump field at $\omega_{0}$ can then combine with the field at $\omega_{3}$ in a sum frequency generation process~\cite{SFG}, to produce a further field at $\omega_{2}$, with effective nonlinearity $\kappa_{2}$. In Yu \etal~\cite{Yu} the nonlinear medium is a quasiperiodic superlattice.  We will use the annihilation operators $\hat{a}_{j}$ to describe the fields at $\omega_{j}$ for $j=1,2,3$. If we consider that the pump field is intense and classical so that depletion does not become important, we may write the interaction Hamiltonian as
\begin{equation}\label{eq:Hint}
{\cal H}_{int} =
i\hbar\kappa_{1}(\hat{a}_1^\dag\hat{a}_3^\dag-\hat{a}_1\hat{a}_3)+i\hbar\kappa_{2}(\hat{a}_3\hat{a}_2^\dag-\hat{a}_3^\dag\hat{a}_2).
\label{eq:CPham}
\end{equation}
In this case, the $\kappa_{j}$ represent products of the actual nonlinearity multiplied by the amplitude of the pump field.

We find the Heisenberg equations of motion for the annihilation operators as
\begin{eqnarray}
\frac{d\hat{a}_{1}}{dt} &=& \kappa_{1}\hat{a}_{3}^{\dag}, \nonumber \\
\frac{d\hat{a}_{2}}{dt} &=& \kappa_{2}\hat{a}_{3}, \nonumber \\
\frac{d\hat{a}_{3}}{dt} &=& \kappa_{1}\hat{a}_{1}^{\dag}-\kappa_{2}\hat{a}_{2}, 
\label{eq:CPErnie}
\end{eqnarray}
with those for the creation operators being the Hermitian conjugates.
These may be solved analytically. Setting $\zeta = \sqrt{\kappa_{1}^{2}-\kappa_{2}^{2}}$ for $\kappa_{1}>\kappa_{2}$, we find
\begin{eqnarray}
\hat{a}_{1}(t) &=& \frac{\kappa_{1}^{2}\cosh \zeta t-\kappa_{2}^{2}}{\zeta^{2}}\hat{a}_{1}(0)-\frac{\kappa_{1}\kappa_{2}(\cosh\zeta t-1)}{\zeta^{2}}\hat{a}_{2}^{\dag}(0)+\frac{\kappa_{1}\sinh\zeta t}{\zeta^{2}}\hat{a}_{3}^{\dag}(0), \nonumber \\
\hat{a}_{2}(t) &=&\frac{\kappa_{1}\kappa_{2}(\cosh\zeta t-1)}{\zeta^{2}}\hat{a}_{1}^{\dag}(0) + \frac{\kappa_{1}^{2}-\kappa_{2}^{2}\cosh\zeta t}{\zeta^{2}}\hat{a}_{2}(0)+\frac{\kappa_{1}\sinh\zeta t}{\zeta^{2}}\hat{a}_{3}(0), \nonumber \\
\hat{a}_{3}(t) &=& \frac{\kappa_{1}\sinh\zeta t}{\zeta^{2}}\hat{a}_{1}^{\dag}(0)-\frac{\kappa_{2}\sinh\zeta t}{\zeta^{2}}\hat{a}_{2}(0)+\cosh\zeta t\;\hat{a}_{3}(0).
\label{eq:CPsols}
\end{eqnarray}
We note here that these are different to the solutions given by Ferraro \etal~\cite{Ferraro}, who worked in the regime where $\kappa_{2}>\kappa_{1}$. They have been given previously by Olsen and Bradley~\cite{OlsenBradley}, who also calculated the non-optimised VLF measures of Eq.~\ref{eq:VLF}, but did not investigate bipartite entanglement.
This immediately allows us to write solutions for the quadrature operators,
which then allows us to find expressions for all the entanglement and EPR-steering correlations of section \ref{sec:inequalities}. Setting
\begin{eqnarray}
\alpha &=& \frac{\kappa_{1}^{2}\cosh \zeta t-\kappa_{2}^{2}}{\zeta^{2}}, \nonumber \\
\beta &=& \frac{\kappa_{1}\kappa_{2}(\cosh\zeta t-1)}{\zeta^{2}}, \nonumber \\
\gamma &=& \frac{\kappa_{1}\sinh\zeta t}{\zeta^{2}}, \nonumber \\
\delta &=& \frac{\kappa_{1}^{2}-\kappa_{2}^{2}\cosh\zeta t}{\zeta^{2}}, \nonumber \\
\epsilon &=& \frac{\kappa_{2}\sinh\zeta t}{\zeta^{2}}, \nonumber \\
\eta &=& \cosh\zeta t,
\label{eq:Greek}
\end{eqnarray}
we find the moments required for the variances and covariances as
\begin{eqnarray}
\langle \hat{X}_{1}^{2}\rangle &=& \langle \hat{Y}_{1}^{2}\rangle = \alpha^{2}+\beta^{2}+\gamma^{2}, \nonumber \\
\langle \hat{X}_{2}^{2}\rangle &=&  \langle \hat{Y}_{2}^{2}\rangle =  \beta^{2}+\delta^{2}+\gamma^{2}, \nonumber \\
\langle \hat{X}_{3}^{2}\rangle &=&  \langle \hat{Y}_{3}^{2}\rangle =  \gamma^{2}+\epsilon^{2}+\eta^{2}, \nonumber \\
\langle \hat{X}_{1}\hat{X}_{2}\rangle &=& \alpha\beta-\beta\delta+\gamma^{2}, \nonumber \\
\langle \hat{X}_{1}\hat{X}_{3}\rangle &=& \alpha\gamma+\beta\epsilon+\gamma\eta, \nonumber \\
\langle \hat{X}_{2}\hat{X}_{3}\rangle &=& \gamma\beta-\delta\epsilon+\gamma\eta, \nonumber \\
\langle \hat{Y}_{1}\hat{Y}_{2}\rangle &=& -\alpha\beta+\beta\delta-\gamma^{2}, \nonumber \\
\langle \hat{Y}_{1}\hat{Y}_{3}\rangle &=& -\alpha\gamma-\beta\epsilon-\gamma\eta, \nonumber \\
\langle \hat{Y}_{2}\hat{Y}_{3}\rangle &=& \beta\gamma-\delta\epsilon+\gamma\eta.
\label{eq:CPXYmoments}
\end{eqnarray}

\begin{figure}[tbp]
\includegraphics[width=0.75\columnwidth]{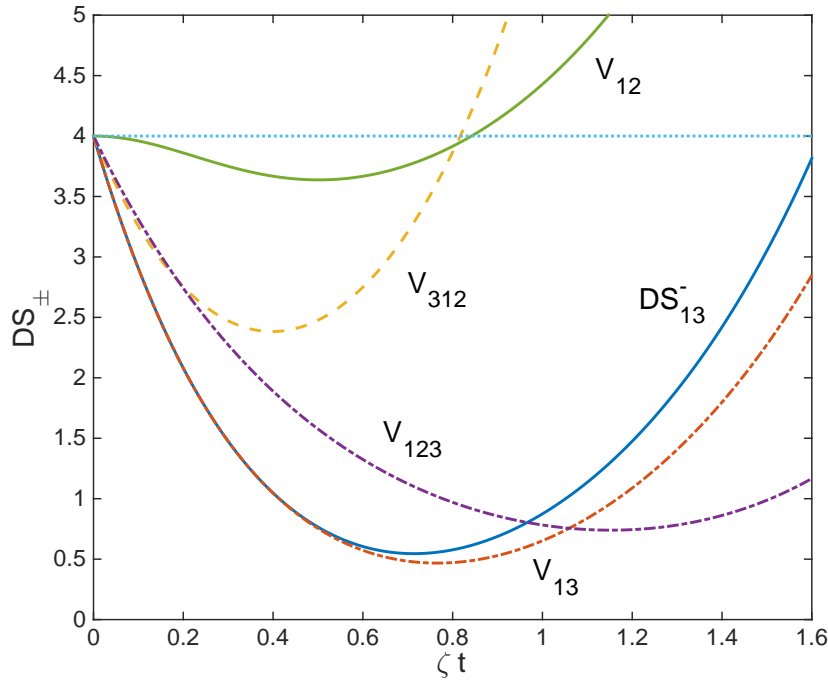}
\caption{(colour online) DS$_{13}^{-}$, V$_{123}$, V$_{312}$, V$_{12}$, and V$_{13}$ for the asymmetric model, with $\kappa_{2}=0.6\kappa_{1}$ and $\kappa_{1}=1$. We see that both bipartite and tripartite entanglement are predicted over a range of interaction strength $\zeta t$. Note that the line at $4$ is a guide to the eye.}
\label{fig:Clervie}
\end{figure}

\begin{figure}[tbp]
\includegraphics[width=0.75\columnwidth]{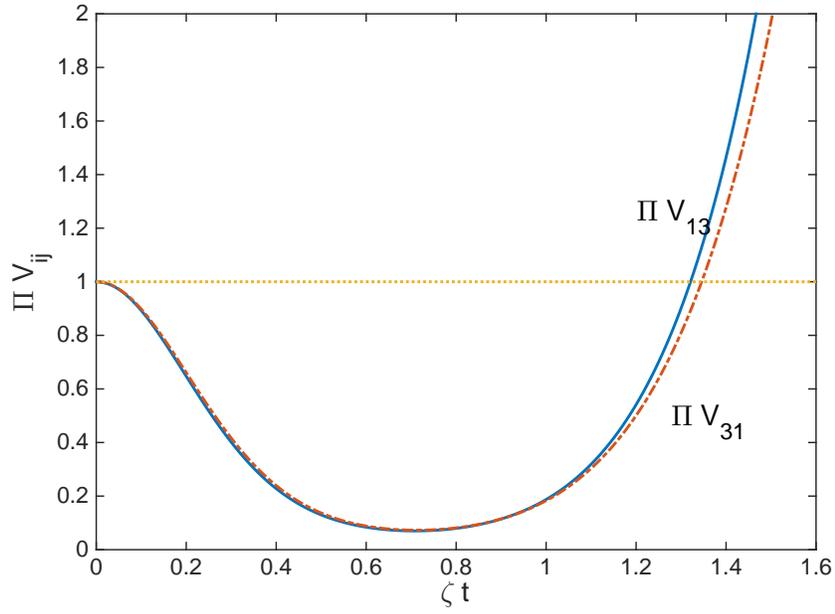}
\caption{(colour online) The EPR-steering correlations for modes $1$ and $3$ of the asymmetric model, with $\kappa_{2}=0.6\kappa_{1}$ and $\kappa_{1}=1$. Note that the line at $1$ is a guide to the eye.}
\label{fig:EPRClervie}
\end{figure}

In Fig.~\ref{fig:Clervie} we show that both tripartite and bipartite entanglement are predicted over a range of interaction strength, $\zeta t$, for $\kappa_{2}=0.6\kappa_{1}$. The Duan-Simon measure, DS$_{13}^{-}$, shows that modes $1$ and $3$ are entangled, while either of V$_{312}$ or V$_{123}$, in the region where they are less than $4$, demonstrate tripartite entanglement. Fig.~\ref{fig:EPRClervie} shows the bipartite EPR-steering correlations between modes $1$ and $3$, demonstrating that these two modes are able to steer each other. There was no violation of either the bipartite entanglement or EPR-steering inequalities for the pairs $1,2$ and $2,3$.

\begin{figure}[tbp]
\includegraphics[width=0.75\columnwidth]{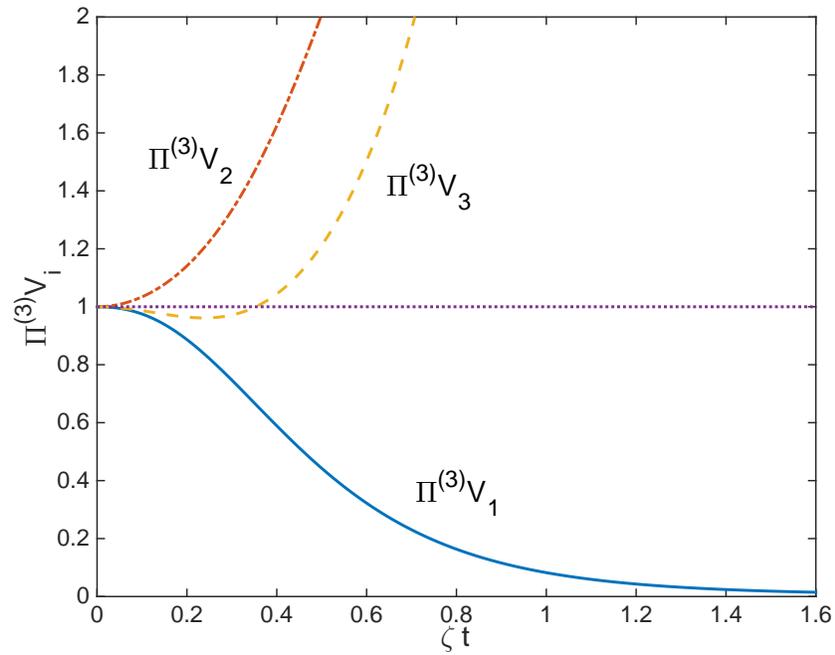}
\caption{(colour online) The EPR-steering correlations $\Pi^{(3)}V_{i}$ for the asymmetric model, with $\kappa_{2}=0.6\kappa_{1}$ and $\kappa_{1}=1$. Note that the line at $1$ is a guide to the eye.}
\label{fig:CEPR3Vi}
\end{figure}

\begin{figure}[tbp]
\includegraphics[width=0.75\columnwidth]{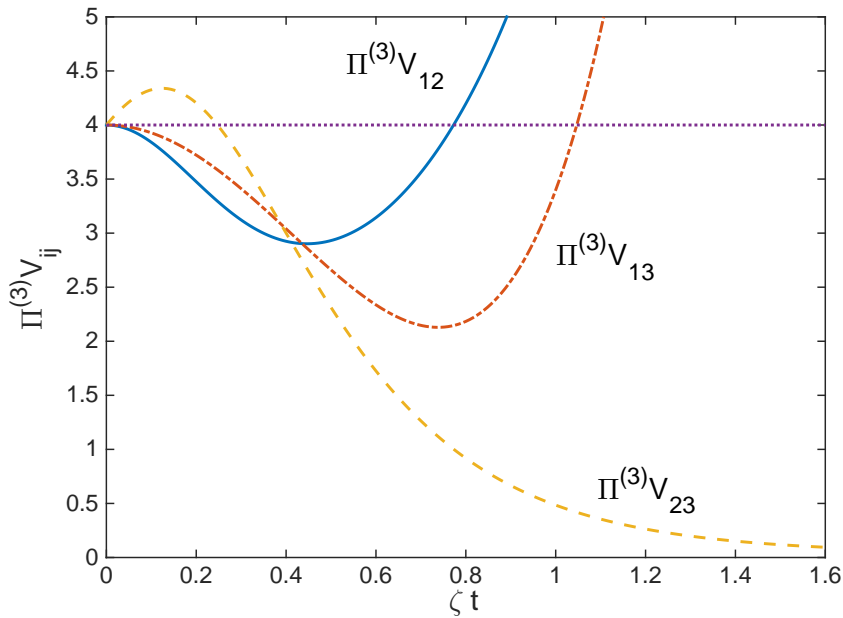}
\caption{(colour online) The EPR-steering correlations $\Pi^{(3)}V_{ij}$ for the asymmetric model, with $\kappa_{2}=0.6\kappa_{1}$ and $\kappa_{1}=1$. Note that the line at $4$ is a guide to the eye.}
\label{fig:CEPR3Vij}
\end{figure}

When we investigate the three-mode EPR-steering correlations, shown in Fig.~\ref{fig:CEPR3Vi} and Fig.~\ref{fig:CEPR3Vij}, we see that only $\Pi^{(3)}V_{2}$ fails to violate the inequality, although the violation by $\Pi^{(3)}V_{3}$ is minimal. This means that, for these parameters, modes $1$ and $3$ cannot be used to steer mode $2$, although all the other combinations are possible over some range of interaction strengths. This is quite different from the symmetric case in the previous section, where the correlations were equivalent under any change of indices. We also draw attention to the fact that the system does not enter the T state regime within the range of interaction strength shown here, although it will for longer interaction times.

\section{Conclusions}

We have investigated two symmetric and one asymmetric systems known to produce tripartite entangled outputs, in terms of both the bipartite and tripartite entanglement available. We found that all three systems produce both tripartite and bipartite entanglement in some operating regimes. In these regimes the outputs may be thought of as W-type states, even though they do not satisfy the criteria completely. Only the asymmetric system was found to produce bipartite EPR-steering. All three systems have operating regimes where only tripartite entangled outputs are introduced, where the outputs qualify as T states. In the limit of large interaction strength or squeezing, the two symmetric systems produce states which may truly be thought of as having GHZ properties. However, this limit is unphysical because either the undepleted pump approximation or energy conservation breaks down long before it is reached. Although the labelling of these states as GHZ or GHZ type is common, some care should be taken with this since they do not satisfy all the criteria. For example, in the physically attainable regimes they do not produce eigenstates of quadrature combinations and thus will not give a yes or no reply to the question of whether tripartite entanglement is present. The answer they give arises from a statistical violation of the entanglement criteria. 

 On a final note, the production of qualitatively different quantum states from the same apparatus, obtained by changing the operating parameters, may be of advantage to quantum information experimentalists. There may be situations where changing the input laser intensities, for example, and moving from a W type state to a T state is advantageous. As was shown in Ref.~\cite{1SQKD}, there are operating regimes of the asymmetric system where two of the participants can practise one-sided device independent quantum key distribution which cannot involve the third. We expect that there will be other applications which take advantage of the flexibility we have demonstrated in this article.

\section*{Acknowledgments}

This research was supported by the Australian Research Council under the Future Fellowships Program (Grant ID: FT100100515).


\end{document}